\documentclass{iopjournal}

\usepackage{graphicx}
\usepackage{epsfig}
\usepackage{epstopdf}
\usepackage{amsmath}
\usepackage{amssymb}
\usepackage{ragged2e}

\begin{document}

\articletype{Paper}

\title{A Machine-Learning Based Approach to the Evaluation of the Critical Scaling Behavior of Anisotropic Spin Systems}

\author{Alina A. Chubarova$^{1,*}$\orcid{0009-0009-0414-1963}, Ivan A. Mamonov$^2$\orcid{0000-0002-6522-2873}, Marina V. Mamonova$^2$\orcid{0000-0001-7466-086X}, Mikhail I. Bogachev$^3$\orcid{0000-0002-0356-5651} and Pavel V. Prudnikov$^1$\orcid{0000-0002-6522-2873}}

\affil{$^1$Center of New Chemical Technologies BIC, Boreskov Institute of Catalysis, Neftezavodskaya st. 54, Omsk 644040, Russia}

\affil{$^2$Theoretical Physics Department, Dostoevsky Omsk State University, Mira av. 55A, Omsk 644077, Russia}

\affil{$^3$Faculty of Radio Engineering \& Telecommunication, St.~Petersburg State Electrotechnical University ``LETI'', 5 Professor Popov street, St.~Petersburg 197022, Russia}

\affil{$^*$Author to whom any correspondence should be addressed.}

\email{alina\_chubarova@ihcp.ru}


\begin{abstract}
\justify
Computational models adequately representing phase transitions and evaluating the critical system parameters are essential for the understanding of the properties of a wide range of materials. Here we propose a machine learning (ML)-based approach to the identification of the critical point in anisotropic spin systems. Our approach implies training of a convolutional neural network (CNN) model from the correlation matrices obtained by Monte Carlo simulations. Next, the pretrained model is employed as a fast estimator of the critical temperature, which can be extracted in several complementary ways from the CNN model inference, this way improving the robustness of the analysis. The ML-based estimates obtained in this study are in very good agreement with the reference Monte Carlo simulation results, while computational costs are about 10x lower compared to the classical thermodynamic approach.
\end{abstract}

\justify

\section{Introduction}
Anomalous fluctuations play a crucial role in critical phenomena. For three-dimensional systems, it is generally impossible to obtain exact predictions for all relevant quantities from a realistic microscopic model. In many cases, only the scaling hypothesis makes it possible to describe universal critical behavior through \textit{data collapse} \cite{Cardy96}. Thus, numerical computational models often represent the only available approach, while direct numerical simulations using the conventional Monte Carlo approach require substantial computational resources due to the critical slowing-down effect in the vicinity of the critical temperature.

Machine Learning (ML) algorithms have demonstrated considerable potential for the analysis of statistical and condensed-matter systems. Their importance is not limited to replacing conventional numerical techniques; rather, they provide an additional way to extract physically meaningful information from large ensembles of spin configurations, correlation functions, or other high-dimensional simulation data. In the context of critical phenomena, this is especially important because the vicinity of the critical point is characterized by strong fluctuations, finite-size effects, and long relaxation times. Therefore, a central question is whether machine-learning outputs can be used not only for phase classification, but also for quantitative estimates of critical parameters.

Several recent studies have addressed this question from different perspectives. Chertenkov et al. \cite{Shchur} proposed a neural-network-based finite-size analysis for critical phenomena, showing that neural-network classification outputs can be connected with the methodology of finite-size scaling. This is important because it goes beyond a purely qualitative identification of phases.
However, further tests are required for anisotropic three-dimensional systems, where crossover effects and finite-size corrections may influence the scaling behavior. Tanaka and Tomiya \cite{Tanaka} demonstrated that convolutional neural network (CNN) can detect phase transitions directly from spin configurations, while Carrasquilla and Melko \cite{Carrasquilla} showed that supervised neural networks can efficiently classify phases of matter. These studies established neural networks as powerful phase-recognition tools, but the use of network-derived probabilistic observables for quantitative finite-size scaling analysis remains less developed.

ML studies of spin systems have also revealed several important methodological limitations. Shiina et al. \cite{Shiina} investigated spin models using ML approaches and demonstrated that neural networks can distinguish different phase regions, including nontrivial cases associated with clock-model behavior. This illustrates the sensitivity of ML to features beyond simple magnetization-based order. At the same time, such studies also show that the interpretation of learned features requires care, especially when the phase structure is more complex than in the standard Ising model. Azizi and Pleimling \cite{Azizi} further emphasized this point by investigating ML--generated configurations in systems with conserved quantities and demonstrating that generative approaches may produce physically inconsistent states if conservation laws are not properly respected. Thus, ML methods must be combined with physically motivated input data and conventional statistical checks.

Unsupervised and weakly supervised approaches form another important direction. Wang \cite{Wang_L} showed that phase transitions can be identified without prior knowledge of the order parameter. Wetzel \cite{Wetzel} compared principal component analysis with variational autoencoders for unsupervised detection of phase transitions. Wang and Zhai \cite{Wang_C,Wang_C2} applied principal component analysis and kernel principal component analysis to frustrated classical spin models, whereas Hu et al. \cite{Hu} critically examined the ability of unsupervised learning methods to distinguish phases, phase transitions, and crossovers. Ponte and Melko \cite{Ponte} demonstrated that kernel methods can provide an interpretable route to identifying order parameters, and Zhang et al. \cite{Zhang} introduced a few-shot learning strategy for recognizing phase transitions from a limited number of training samples. These studies are important because they reduce the need for manually specified order parameters. However, they are often focused on qualitative phase identification or benchmark models, whereas the extraction of critical temperatures and scaling behavior from finite-size data remains a separate problem.

Other works have extended data-driven and computational approaches to systems with more complex spin textures and phase diagrams. Wang et al. \cite{Wang} demonstrated that neural networks can learn order parameters from videos of skyrmion dynamical phases, while Mazurenko et al. \cite{Mazurenko} considered the problem of estimating patterns of classical and quantum skyrmion states. G\'{o}mez Albarrac\'{\i}n and Rosales \cite{Gomez} used ML techniques to construct detailed phase diagrams for skyrmion systems. These works are important because they show that modern numerical and ML approaches can operate beyond simple equilibrium benchmark models and can be used to analyze complex magnetic textures, including topologically nontrivial spin structures. However, such applications are primarily aimed at phase recognition, pattern estimation, or phase-diagram reconstruction. They do not directly address the problem of extracting finite-size scaling information in anisotropic three-dimensional Heisenberg systems.

The need for efficient numerical approaches is also evident in complex disordered spin systems. Kumar et al. \cite{Kumar} investigated the critical behavior of the three-dimensional random-field Potts model using a quasi-exact ground-state algorithm. Although this work is not an ML study, it is relevant here because it illustrates the computational difficulty of determining critical behavior in complex three-dimensional spin models. This supports the broader motivation for developing complementary numerical tools that can reduce the computational effort required for critical-parameter estimation.

Generative and probabilistic ML models have also been employed to study statistical systems. Torlai and Melko \cite{Torlai} demonstrated that Boltzmann machines can learn thermodynamic properties of statistical systems, whereas Morningstar and Melko \cite{Morningstar} applied deep learning to the Ising model in the vicinity of criticality. Rao et al. \cite{Rao} used restricted Boltzmann machines to identify product-order phases in quantum systems. These approaches are valuable because they attempt to learn statistical distributions rather than only classify configurations. However, for critical-point estimation one must still ensure that the learned representation is physically meaningful and that the resulting quantities can be compared with conventional finite-size scaling observables.

Neural-network-based phase-recognition methods have been tested on a broad range of models. The confusion scheme proposed by van Nieuwenburg et al. \cite{Nieuwenburg} provides a way to locate phase transitions without a known critical point. Kim and Kim \cite{Kim} investigated the minimal neural-network architecture required to reproduce Ising criticality. Zhang et al. \cite{Zhang_W} extended ML analysis to percolation and XY models, Li et al. \cite{Li} applied neural networks to two-dimensional Potts models, and Beach et al. \cite{Beach} demonstrated that ML can identify vortex structures associated with the Kosterlitz--Thouless transition. These works are important benchmarks because they show that neural networks can detect different types of critical behavior. Nevertheless, many of these studies focus on two-dimensional or well-established model systems, whereas anisotropic three-dimensional spin models require additional analysis due to the interplay between anisotropy, finite-size effects, and crossover behavior.

ML techniques have also been extended beyond classical spin systems. Kieferov\'{a} and Wiebe \cite{Kieferová} introduced quantum Boltzmann machines for quantum-state tomography, while Benedetti et al. \cite{Benedetti,Benedetti_1} developed quantum-assisted probabilistic graphical models and Helmholtz machines for hybrid quantum-classical learning. These studies demonstrate the broader relevance of ML for physical systems with high-dimensional state spaces. However, their objectives differ from the present problem, which is focused on the estimation of critical parameters in a classical anisotropic spin model.

Data-driven and high-throughput methods have likewise become important tools in materials science. Curtarolo et al. \cite{Curtarolo} pioneered the use of data mining for crystal-structure prediction from quantum-mechanical calculations, while Morgan et al. \cite{Morgan} discussed high-throughput and data-mining strategies based on ab initio methods. Fischer et al. \cite{Fischer} further combined data-mining approaches with quantum-mechanical calculations for crystal-structure prediction. As an example of computational analysis of complex alloys, Uporov et al. \cite{Rultsev} combined experimental measurements with ab initio calculations to study pressure effects on the electronic structure and electrical conductivity of a TiZrHfNb high-entropy alloy. These studies show the importance of computational and data-driven approaches in materials physics, but they do not directly address the finite-size scaling of magnetic phase transitions.

The two-dimensional Ising model provides a convenient benchmark for tuning ML approaches because its exact solution is known \cite{Shchur,Ising_NN}. However, magnetic materials are often described by more complex spin models for which exact analytical solutions are unavailable. The Heisenberg spin model remains a central framework for understanding magnetic ordering and critical phenomena in real materials. In particular, easy-axis and easy-plane anisotropies, spin--orbit coupling, local dipolar interactions, and dimensional effects can substantially influence correlation lengths, excitation spectra, and renormalization-group flows. Recent studies of magnetic critical behavior in van der Waals magnets and anisotropic magnetic systems confirm the importance of anisotropy and dimensionality for the interpretation of critical behavior \cite{Plumley,Liang,Itou}. Therefore, anisotropic variants of the Heisenberg model are directly relevant for understanding magnetic phase transitions in realistic systems.

Recent studies have also extended phase-recognition ideas to quantum and two-dimensional systems. Sander et al. \cite{Sander} constructed a quantum CNN for phase recognition in two dimensions, Civitcioglu et al. \cite{Civitcioglu} compared phase-determination strategies with and without deep learning, and Abuali et al. \cite{Abuali} showed that deep learning can detect phase transitions even from minimal training examples. These studies highlight the flexibility of neural-network approaches, but they also show that performance depends on the architecture, the training protocol, the choice of input representation, the temperature grid, and the system size. These issues are addressed here for an easy-axis three-dimensional Heisenberg model, where the combination of exchange anisotropy, finite-size effects, and critical slowing down makes the extraction of critical parameters especially demanding.

In the present work, we focus on the three-dimensional anisotropic Heisenberg model with easy-axis exchange anisotropy. The main motivation is to test whether a CNN can be used not only as a phase classifier, but also as a tool for critical-temperature estimation and finite-size scaling analysis. In contrast to approaches based directly on spin configurations or manually selected scalar descriptors, the CNN is trained on correlation matrices obtained from Monte Carlo simulations. This choice allows the network to analyze spatial correlation patterns associated with the transition region. The resulting probabilistic CNN response is then used to estimate the transition temperature and to construct finite-size scaling observables, which are compared with conventional Monte Carlo-based estimates.

The paper is organized as follows. In Sec.~\ref{sec:MM}, we describe the model and the key elements of the machine-learning approach. Different algorithms to estimate a critical temperature $T_c$ for a fixed value of the anisotropy parameter are discussed in Sec.~\ref{sec:crit}. Simulation results and scaling analyses are provided in Sec.~\ref{sec:DataAndScaling}. Finally, an overall summary, conclusions, and outlook are presented in Sec.~\ref{sec:concl}.

\section{Model and methods}\label{sec:MM}
The Hamiltonian of the Heisenberg model with easy-axis anisotropy is given by \cite{Anis_Heis_1}:
\begin{equation}\label{eq:Ham_Heis}
	H = -J \sum \limits_{\langle i,j \rangle}^N [(1-\Delta)(S_i^x S_j^x + S_i^y S_j^y) + S_i^z S_j^z],
\end{equation}
where $J>0$ is the ferromagnetic exchange interaction constant; $S_i^x$, $S_i^y$, and $S_i^z$ are the components of the three-dimensional classical spin vector $\mathbf{S}_i$, located at the $i-$th lattice site; $N=L^3$ is the total number of spins; and $L$ is the linear lattice size. The summation is performed over nearest-neighbor pairs${\langle i,j \rangle}$. In the present study, the exchange anisotropy parameter was fixed at $\Delta = 0.63$ \cite{Anis_Heis_2}.

The spin configurations were generated by Monte Carlo simulations of the 3D classical anisotropic Heisenberg model using the Metropolis algorithm \cite{Metropolis}. The simulations were performed on a simple cubic lattice of size $L \times L \times L$, where $L=8,16,24,32$, and $48$. Periodic boundary conditions were applied in all three spatial directions. The exchange anisotropy parameter was fixed at $\Delta=0.63$. The initial magnetization was $m_0=1.0$, corresponding to an ordered initial state along the easy axis.
	
For each temperature, a single-spin Metropolis update scheme was used. One Monte Carlo sweep consisted of $N=L^3$ attempted spin updates. At each attempted update, a lattice site was selected randomly, a trial spin orientation was generated, and the energy difference $\Delta E$ was calculated. The trial state	was accepted with the Metropolis probability
\begin{equation}\label{W}
	W=\min\left[1,\exp(-\Delta E/T)\right].
\end{equation}
The random numbers were generated using the Mersenne Twister generator. The	temperature interval was $0.0 \leq T/J \leq 3.05$, with a temperature step $\Delta T/J=0.002$. For each temperature, the first $1000$ Monte Carlo sweeps were discarded as equilibration, and the following $2000$ Monte Carlo sweeps were used for data collection. Thus, the total number of Monte Carlo sweeps per temperature was $3000$.
	
The statistical averaging was carried out over $100$ independently generated configurations. The uncertainties were estimated as the root-mean-square deviation of the corresponding quantities with averaging over the configuration ensemble.

To characterize the system across different spatial scales, we used correlation functions calculated from the simulated spin configurations. In the present work, the input observable for the neural-network analysis was the correlation function $C_i(T)$, defined as
\begin{equation}\label{eq:c_heiz}
	C_i(T) = \frac{1}{3({L}/{2})}\sum\limits_{l = {L}/{2}}^L \sum\limits_{\langle j \rangle} \mathbf{S}_{i} \mathbf{S}_{j+l}.
\end{equation}
which reflects the temperature-dependent spatial ordering of the system at distances up to one-half of the lattice size. The values of $C_i(T)$ were obtained at each Monte Carlo step and then used as physically motivated input features for the CNN. Correlation functions are a natural choice in this context because they contain information about the asymptotic critical behavior and encode the spatial structure of fluctuations relevant to the phase transition.
\begin{figure}[h!]
	\centering
	\includegraphics[scale = 1.4]{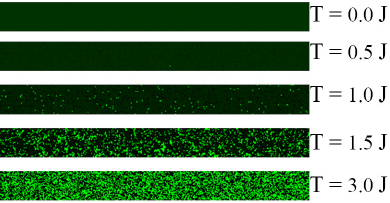}
	\caption{Visualization of correlation matrix at different temperatures.}
	\label{fig:configs}
\end{figure}

The correlation data were arranged as a two-dimensional $L \times L$ matrix for each of the $L$ layers of the system and then converted into image-like arrays for subsequent CNN processing. Figure~\ref{fig:configs} represents an exemplified visualization of correlation matrices at different temperatures, indicating that the correlation patterns differ substantially between the ordered and disordered states, whereas the strongest fluctuations arise in the vicinity of the critical region. This temperature sensitivity provides the physical basis for using CNNs as a phase-classification tool.

The target labels used for CNN training were assigned according to temperature intervals. No threshold value of the correlation function, clustering procedure, or manual feature selection based on the average matrix intensity was used. The  full temperature interval considered in the simulations was $0.0 \leq T/J \leq 3.05$. The ML protocol consisted of two consecutive stages. In the first, coarse-training stage, the separation boundary between the two classes was chosen as $T/J=1.5$, which allowed us to localize	the transition region over the full temperature interval. In the second,	refined-training stage, the critical temperature point was chosen as $T/J=1.645$. Thus, configurations with $0.0 \leq T/J \leq 1.644$ were assigned to the low-temperature class, whereas	configurations with $1.646 \leq T/J \leq 3.05$ were assigned to the high-temperature class. The temperature point $T/J=1.645$ was treated as the separating critical point and was not assigned to either class during the refined labeling procedure.
	
This two-stage procedure was used to obtain a preliminary localization of the transition region and then to improve the accuracy of the critical-temperature estimate in the refined analysis. The CNN analysis was performed using the statistical configurations generated by the Monte Carlo simulations described above. For each lattice size \(L\), a separate CNN was trained from scratch using the same architecture, optimizer, loss function, and training protocol.

The CNN input consisted of the complete correlation matrices arranged as image-like arrays. Therefore, the network did not use manually prescribed scalar descriptors such as the average matrix intensity, the density of nonzero pixels, or a thresholded value of the correlation function. The relevant features were learned automatically by the convolutional filters from the full spatial pattern of the matrices, including local contrast, gradients, extended correlated regions, and their temperature-dependent changes.

\begin{figure}[h!]
	\centering
	\includegraphics[width=0.9\textwidth]{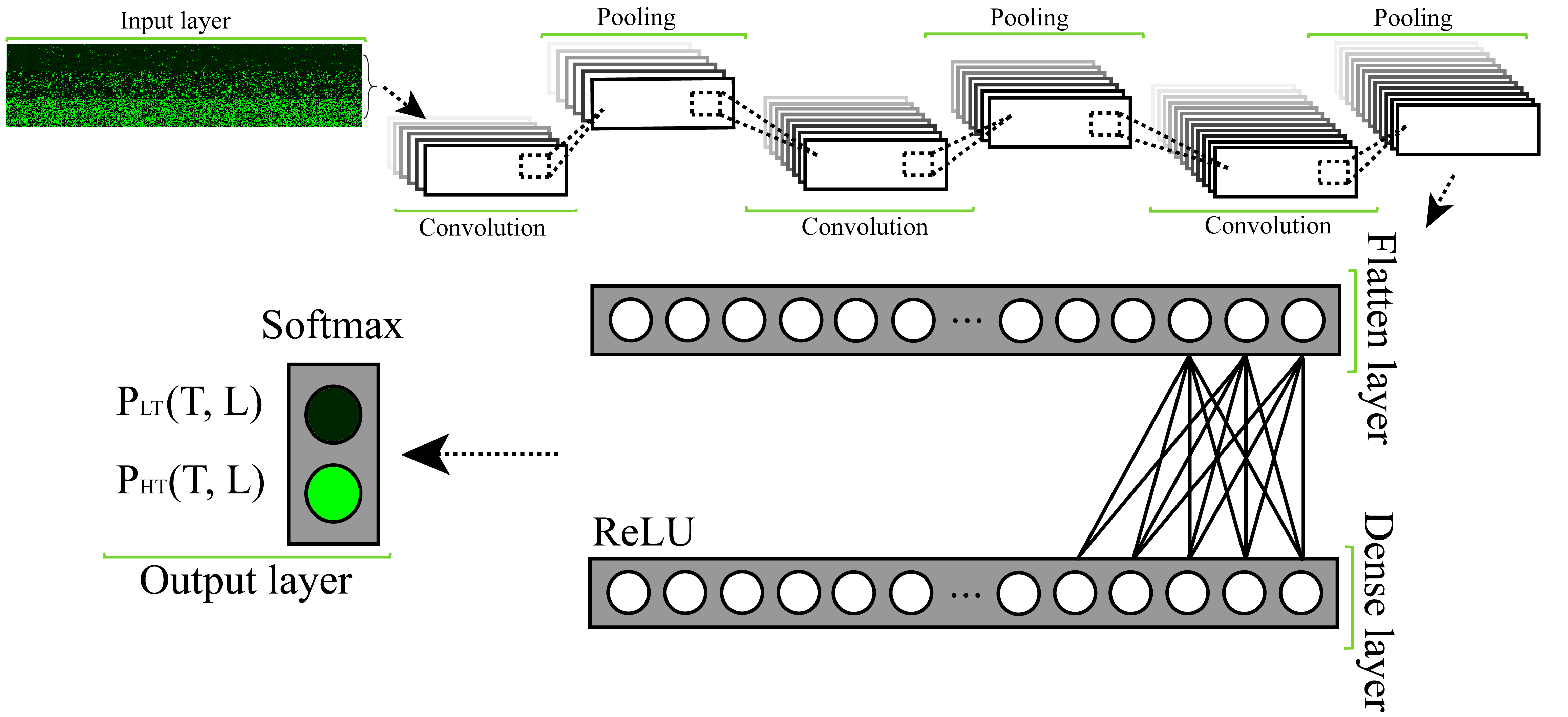}
	\caption{\label{fig:model} The architecture of CNN  employed in this study.}
\end{figure}

A schematic representation of the CNN architecture is shown in Fig.~\ref{fig:model}. The network consists of three convolutional blocks with intermediate max-pooling operations, followed by a flattening stage, a fully connected layer, and a binary Softmax output layer. Thus, Fig.~\ref{fig:model} illustrates the overall workflow, including both feature extraction and subsequent phase classification steps.

Implementation of the above scheme is based on a customization of the TensorFlow implementation \cite{TensorFlow} of a standard lightweight convolutional classifier adapted to the analysis of correlation-matrix patterns. The model was implemented as a sequential CNN with three convolutional layers containing 20, 40, and 80 filters and kernel sizes of $7 \times 7$, $5 \times 5$, and $3 \times 3$, respectively. Each convolutional layer employs ReLU activation and the same padding and is followed by a MaxPooling layer with pool sizes $7 \times 7$, $5 \times 5$, and $3 \times 3$. The extracted feature maps are then flattened and passed to a dense layer with 1024 neurons and ReLU activation, followed by dropout with a rate of 0.2. The output layer is a binary one, corresponding to the low-temperature and high-temperature classes, respectively.

In the implemented training pipeline, the network input shape is $(L, L^2, 3)$. Training was performed using the Adam optimizer and the sparse categorical cross-entropy loss function. The training directory was split into training and validation subsets with a validation fraction of 0.1 and random seed 43, while the test set was loaded separately without shuffling. Classification accuracy was monitored during training only as an auxiliary metric characterizing the quality of phase discrimination.

A separate CNN was trained from scratch for each lattice size $L$. The same network architecture, optimizer, loss function, and training protocol were used for all values of $L$, but the trainable weights were initialized independently and were not shared between different lattice sizes.


Since the CNN was trained as a binary classifier, the critical temperature was not predicted directly by regression. Instead, the binary Softmax output was interpreted as a continuous probabilistic response. For each temperature $T$ and lattice size $L$, the output probabilities corresponding to the low-temperature and high-temperature phases, $\mathrm{P_{\mathbf{LT}}}(T,L)$ and $\mathrm{P_{\mathbf{HT}}}(T,L)$, were averaged over the test configurations
\begin{equation}\label{eq:P}
	P(T,L) = \frac{1}{n_s}\sum_{i = 1}^{n_s}p_i(T,L).
\end{equation}
where $n_s$ is the number of test samples and $p_i(T,L)$ is the output probability for sample $i$. In this way, the network was used not merely for hard phase assignment but for constructing smooth temperature-dependent observables from which the transition point could be inferred.

To quantify fluctuations in the network response, we calculated the deviation function
\begin{equation}\label{eq:Er}
	\mathrm{Er}(T,L) =\frac{1}{n_s} \sum_{i = 1}^{n_s}p_i^2(T,L) - \left(\frac{1}{n_s}\sum_{i = 1}^{n_s}p_i(T,L)\right)^2.
\end{equation}

The resulting $P(T,L)$ and $\mathrm{Er}(T,L)$ curves were next used to obtain critical temperatures. In particular, the critical region was analyzed by the crossing of the probability curves for different lattice sizes and by the peak positions of $\mathrm{Er}(T,L)$, which were approximated in the critical region by Gaussian-like functions. As shown later in the paper, the crossing of $P(T,L)$ provides one estimate of the transition point, while the maxima of $\mathrm{Er}(T,L)$ yield an independent sequence of finite-size critical temperatures.


An additional estimator was obtained from a CNN-derived normalized finite-size scaling observable with correlation-length-like behavior, defined from the network output as
\begin{equation}\label{eq:ksi_net}
	{\varXi}(T,L)/L = -\ln\left (| \langle P^{\infty} - \mathrm{P_{\mathbf{LT}}}(T,L)\rangle | \right),
\end{equation}
where $P^{\infty} = \Theta (T_c - T)$ and $\Theta$ is the Heaviside step function. 

It should be emphasized that $\Xi(T,L)/L$ is not a conventional normalized correlation length calculated directly from the spin--spin correlation function. Instead, it is a CNN-derived finite-size scaling observable constructed from the probabilistic response of the trained neural network. Since the CNN is trained on correlation matrices, which encode spatial spin correlations in the system, the output probability $\mathrm{P}_{\mathbf{LT}}(T,L)$ can be regarded as a smooth finite-size indicator of the low-temperature phase.

In the thermodynamic limit, this probability is expected to approach a step-like function separating the ordered and disordered phases, whereas in finite systems the transition is rounded in the critical region due to finite-size effects. Therefore, $\Xi(T,L)/L$ characterizes the logarithmic deviation of the finite-size CNN response from the ideal step-like thermodynamic-limit behavior. Although this quantity is not identical to the conventional ratio $\xi(T,L)/L$, it demonstrates analogous crossing behavior near the critical point. Motivated by the finite-size scaling analysis of machine-learning observables near phase transitions~\cite{Maskara} and by the use of proxy quantities for finite-size correlations~\cite{Martin}, the temperature dependence of $\Xi(T,L)/L$ was analyzed for different lattice sizes, providing an independent estimate of the transition temperature.
\begin{figure}
	\centering
	\includegraphics[scale=0.35]{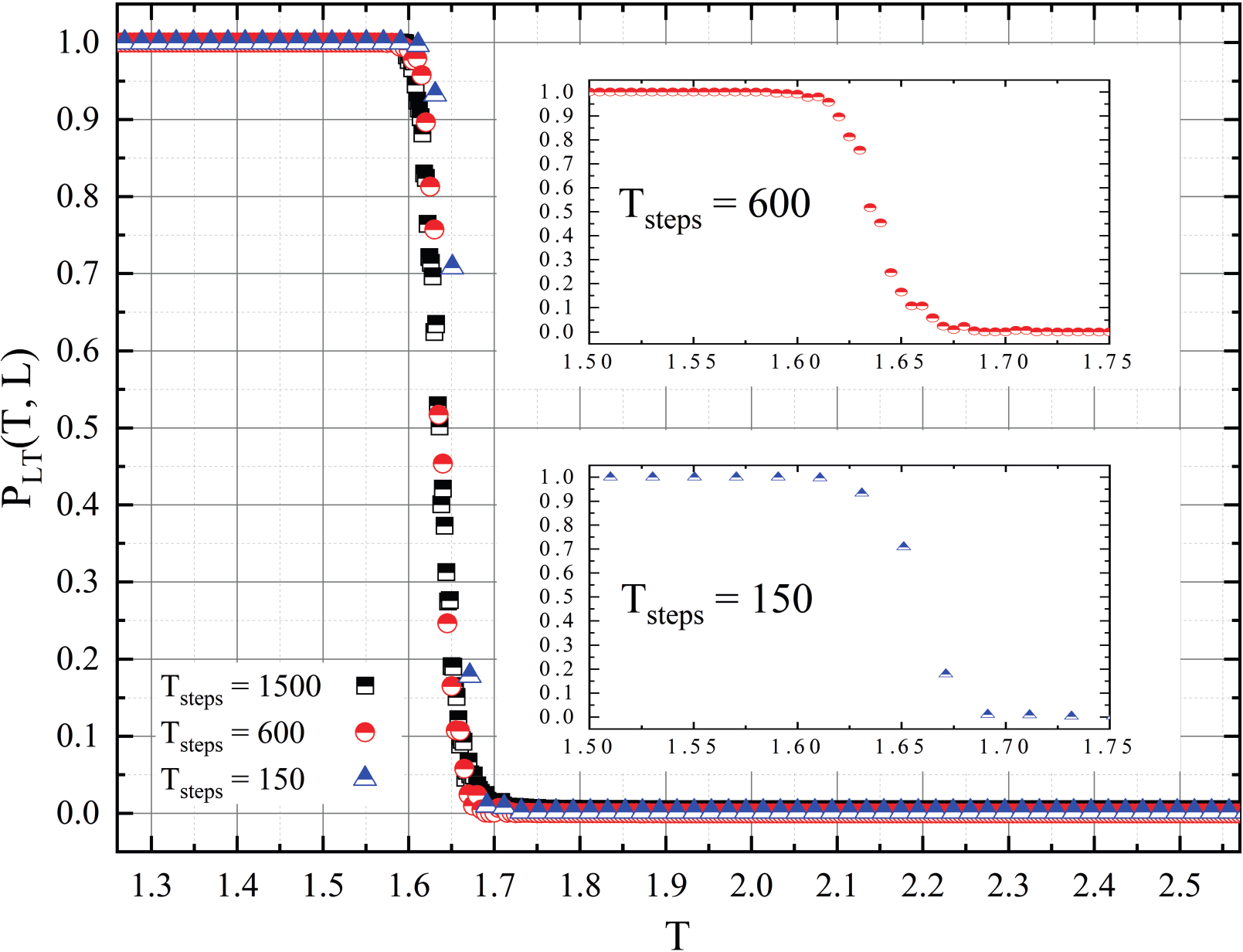}
	\caption{Temperature dependence of $\mathrm{P_{\mathbf{LT}}}$ for $L = 16$. Three temperature grids are compared: $T_{steps}$ = 1500 (squares), $T_{steps}$ = 600 (circles), and $T_{steps}$ = 150 (triangles). The insets magnify the critical region, showing that coarse sampling ($T_{steps}$ = 150) smears the drop near $T_c$, whereas denser grids (600 and 1500) yield nearly identical, sharp transitions.}
	\label{fig:T_steps}
\end{figure}

An important methodological aspect of the present approach is the resolution of the temperature grid. Since the transition point is inferred from the temperature dependence of the CNN output probabilities, the temperature sampling must be sufficiently dense in the critical region. Figure \ref{fig:T_steps} explicitly compares the temperature dependences of $\mathrm{P_{\mathbf{LT}}}$ obtained for three different temperature grids, $T_{steps} =$ 150, 600, and 1500 for $L = 16$. As shown in this figure, coarse temperature discretization ($T_{steps} = 150$) leads to a smeared crossover near $T_c$, whereas denser grids ($T_{steps} = $ 600 and 1500) produce much sharper and nearly identical curves in the transition region. Therefore, reliable determination of the critical temperature requires not only adequate statistical sampling of configurations but also a sufficiently fine discretization of temperature. In this sense, the precision of the method is controlled both by the number of configurations and by the density of the temperature grid used to generate the dataset.

\section{Calculation of critical parameters}\label{sec:crit}
In the following, critical-temperature estimates were obtained from the CNN-based inference and compared against the results of the conventional Monte Carlo simulations to elucidate how the temperature dependence of the phase probabilities extracted from the neural network can be used to determine the location of the critical point of the anisotropic three-dimensional Heisenberg model.
\begin{figure}[h!]
	\centering
	\includegraphics[scale=0.35]{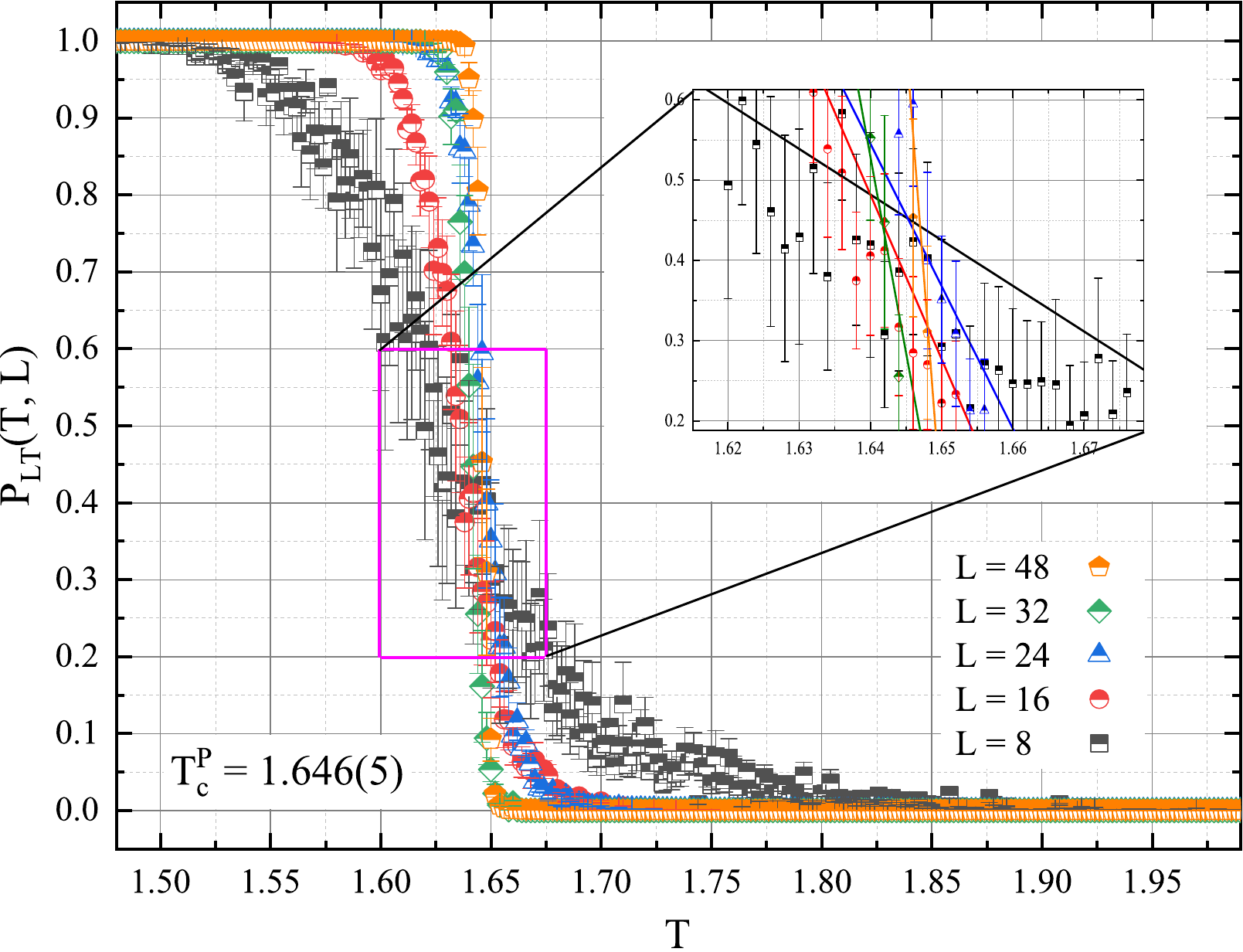}
	\caption{Temperature dependence of $\mathrm{P_{\mathbf{LT}}}$ for $L = 8,16,24,32,48$.}
	\label{fig:pLT_T_all_L}
\end{figure}

Figure~\ref{fig:pLT_T_all_L} shows the temperature dependence of the probability $\mathrm{P_{\mathbf{LT}}}(T,L)$ of the low-temperature phase of the system for lattice sizes $L = 8, 16, 24, 32$ and $48$. As the system approaches the critical region, the probability curves for different lattice sizes exhibit a narrow common intersection region, indicating that the critical temperature can be determined from the CNN inference as $T_{c}^{P} = 1.646(5)$, while near-collapse in the intersection region indicates that the probabilistic response of the neural network provides a stable estimator of the transition point.

For comparison, Ref.~\cite{Prudnikov_Anis_Heis_Tc} reported the value $T_c = 1.6449(5)$ for the three-dimensional Heisenberg model with easy-axis anisotropy using the fourth-order Binder cumulant method. Thus, the value obtained from the CNN probability crossing is in a very good agreement with the reference Monte Carlo estimate.

\begin{figure}[h!]
	\centering
	\includegraphics[width=1.0\textwidth]{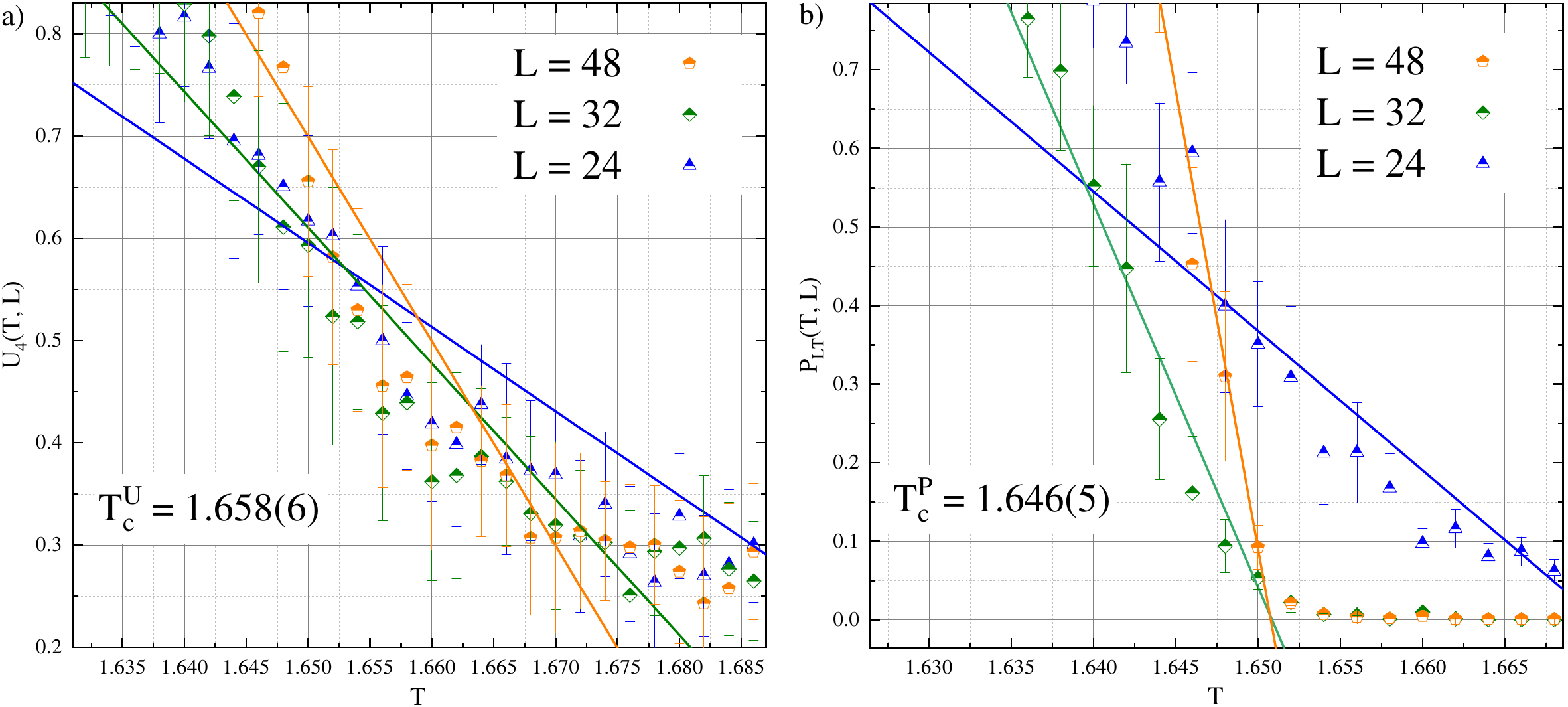}
	\caption{ Determination of the critical temperature of the anisotropic Heisenberg model for linear sizes L = 24, 32, 48 using the methods of a) the 4th order Binder cumulant; b) machine learning.}
	\label{fig:check_ML_MC}
\end{figure}

To compare the machine-learning-based estimate with a conventional thermodynamic approach under the same simulation conditions, we also evaluated the fourth-order Binder cumulant \cite{binder},
\begin{equation}\label{eq:U4}
	U_4(T,L) = \frac{1}{2}\left(3 - \dfrac{ \langle M_4(T,L) \rangle}{ \langle M_2(T,L)\rangle ^2}\right),
\end{equation}
where $M_n(T)$ denotes the $n$-th moment of the magnetization.
\begin{equation}
	M_n = \left\langle \left| \frac{1}{N}\sum_{i=1}^{N}\mathbf{S}_i	\right|^n \right\rangle .
\end{equation}
The corresponding comparison between the Binder-cumulant and CNN-based determinations of the critical temperature is presented in Fig.~\ref{fig:check_ML_MC}.

Figure \ref{fig:check_ML_MC} compares the results obtained for lattice sizes $L=24,32,$ and $48$ using the machine-learning and Monte Carlo approaches under identical simulation conditions. In both cases, the critical temperature was determined from the intersection region of the corresponding curves in the critical domain. The same procedure was applied to the CNN-based inference, specifically to the probability $P_{\mathrm{LT}}(T,L)$ of detecting the system in the low-temperature phase. The analysis indicates that, for the statistics used in the present study, the Binder-cumulant estimate $T_{c}^{U}=1.658(6)$ exceeds both the reference Monte Carlo value $T_c^{MC}=1.6449(5)$~\cite{Prudnikov_Anis_Heis_Tc} and the CNN-based estimate $T_{c}^{P}=1.646(5)$.

Another important difference between the two approaches is the computational time. The time required to calculate the correlation matrix and Binder cumulants under the same simulation parameters differs substantially. In particular, for $L=48$, the calculation of the correlation matrix required approximately 25 hours, whereas the corresponding classical thermodynamic analysis required about 312 hours while using the same computational resources, indicating more than 10-fold benefit.

The above results indicate that the machine-learning-based approach can significantly reduce the computational cost of critical-point estimation for the anisotropic Heisenberg model considered here. At the same time, the accuracy of the method depends strongly on the density of the temperature grid, so a reliable determination of the critical temperature requires sufficiently fine temperature sampling rather than excessively long relaxation and averaging runs.
\begin{figure}[h!]
	\centering
	\includegraphics[scale=0.22]{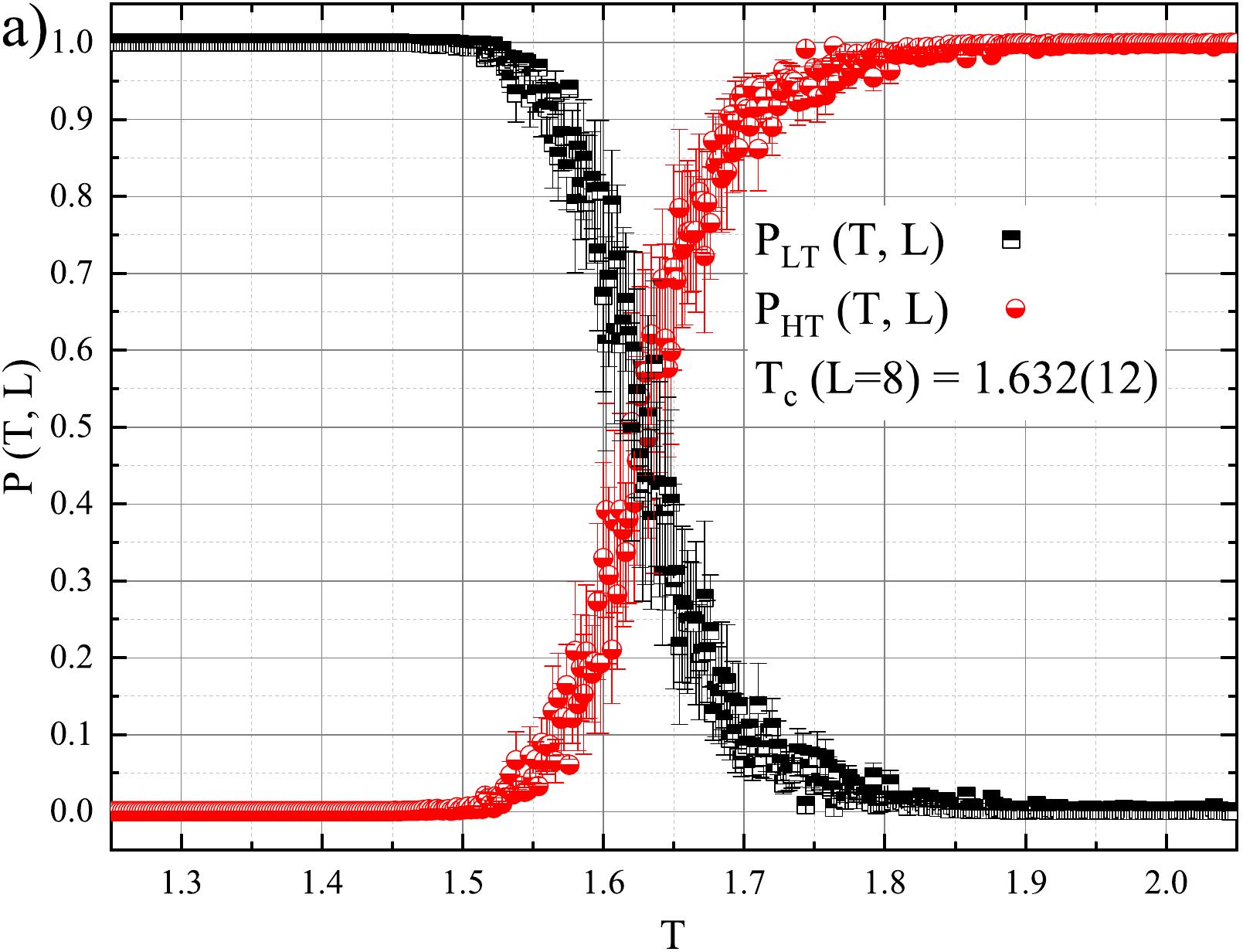}
	\includegraphics[scale=0.22]{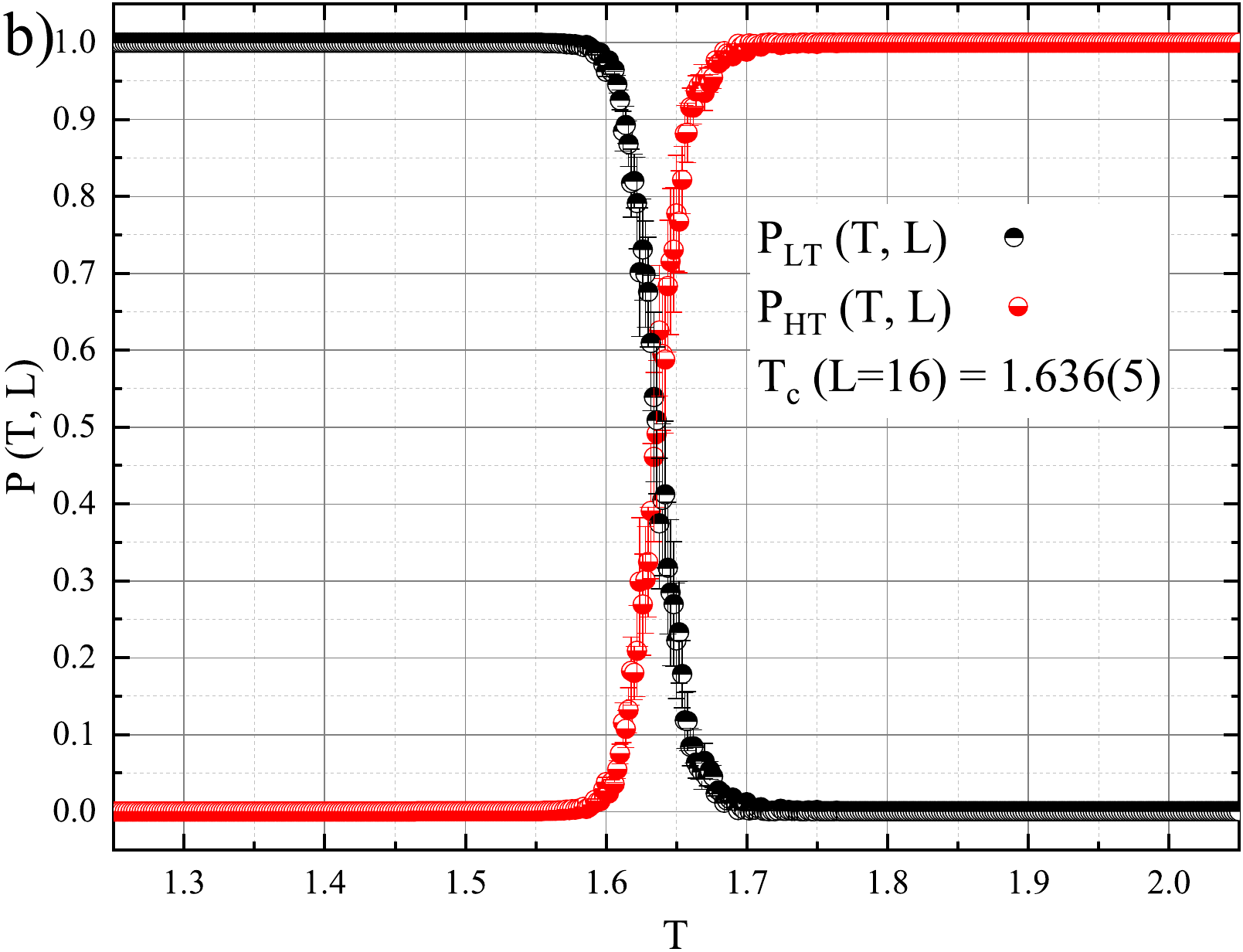}
	\includegraphics[scale=0.22]{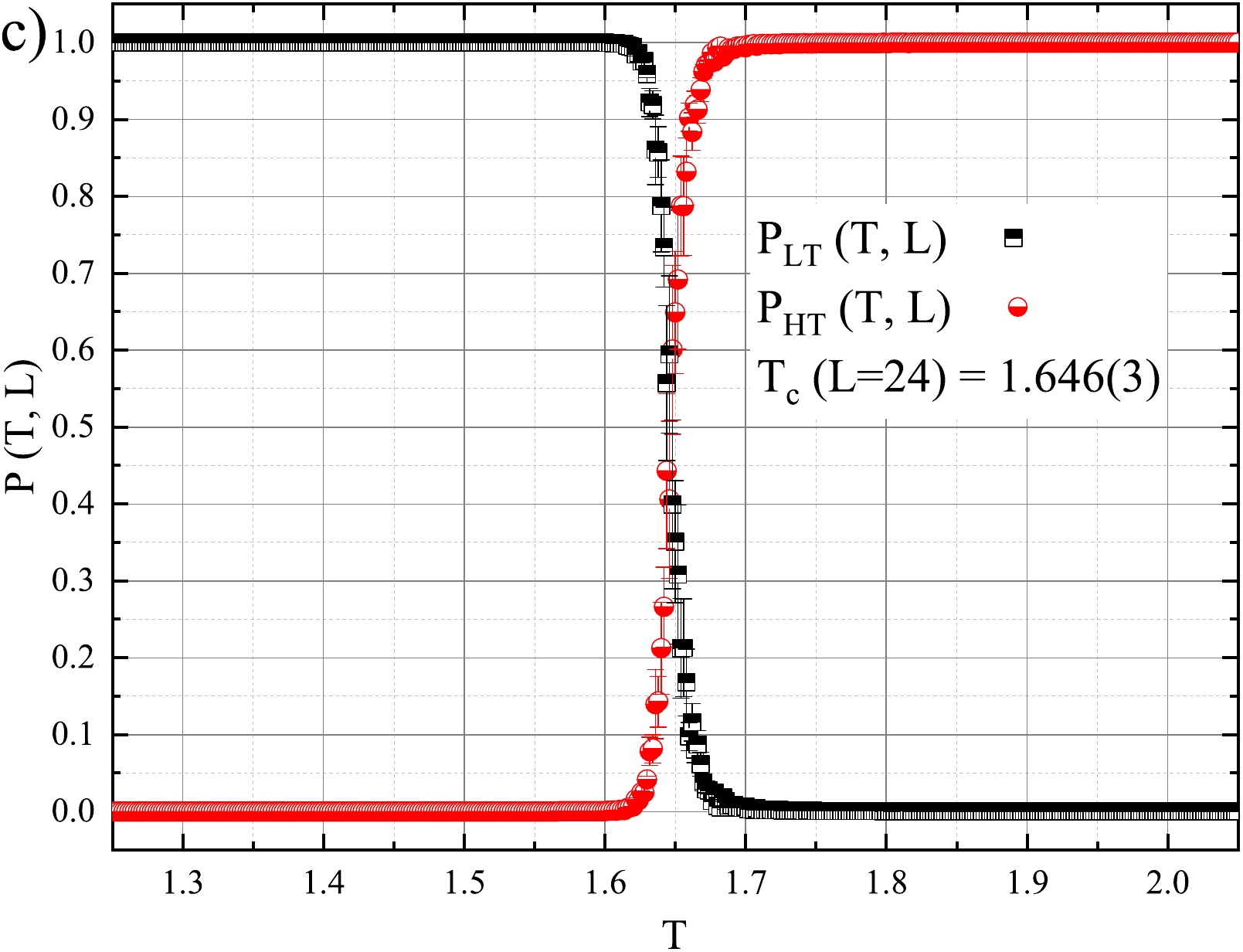}
	\includegraphics[scale=0.22]{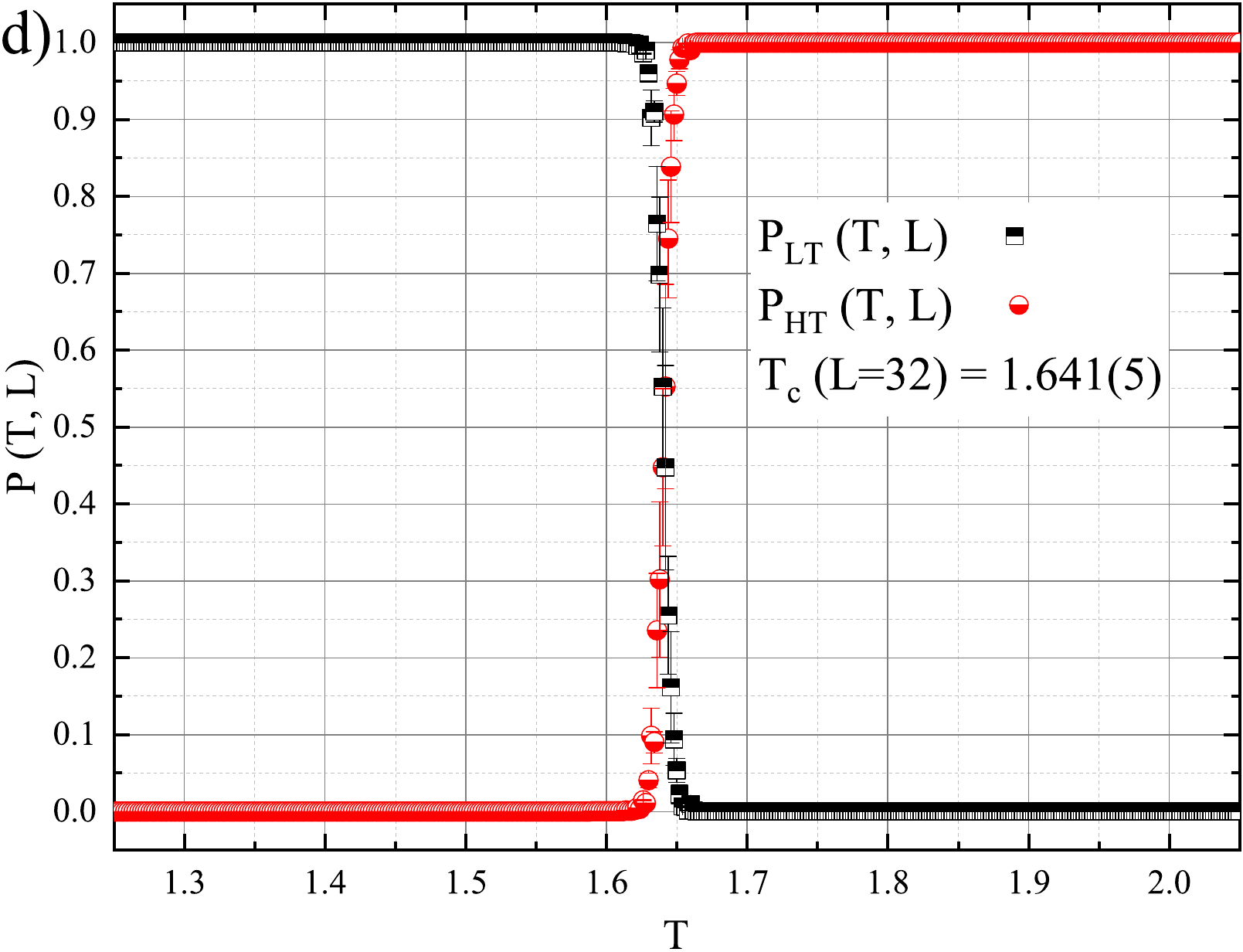}
	\includegraphics[scale=0.22]{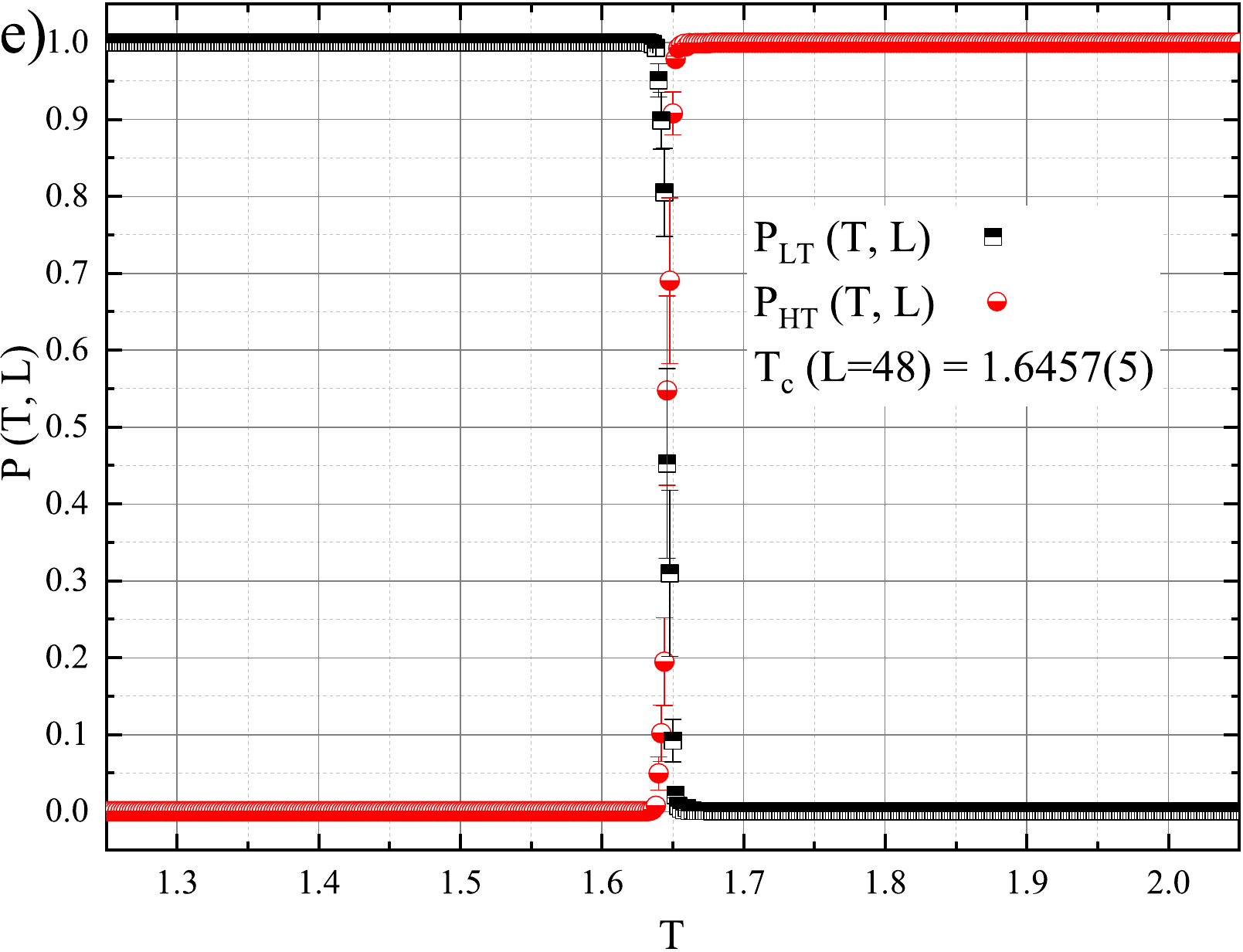}
	\caption{Determination of the critical temperature by the intersection of probabilities for linear sizes: a) $L = 8$; b) $L = 16$; c) $L = 24$; d) $L = 32$; e) $L = 48$.}
	\label{fig:check_P_T}
\end{figure}

Figure \ref{fig:check_P_T} illustrates the determination of the critical temperature from the intersection of the probability curves $\mathrm{P}(T,L)$ for different linear sizes. As the system size increases, the fluctuations in the probability curves become less pronounced, and the neural network yields a sharper and more reliable discrimination between the low- and high-temperature phases.

In \cite{Shchur}, it was proposed to interpret the variance of the CNN-based inference, $\mathrm{P_{LT}}(T,L)$, in terms of the quantity $\mathrm{Er}(T,L)$ defined in Eq.~(\ref{eq:Er}), which captures singular behavior in the Ising and Baxter--Wu models. In the present work, we obtained qualitatively similar behavior for the anisotropic Heisenberg model.
\begin{figure}
	\centering\includegraphics[scale=0.35]{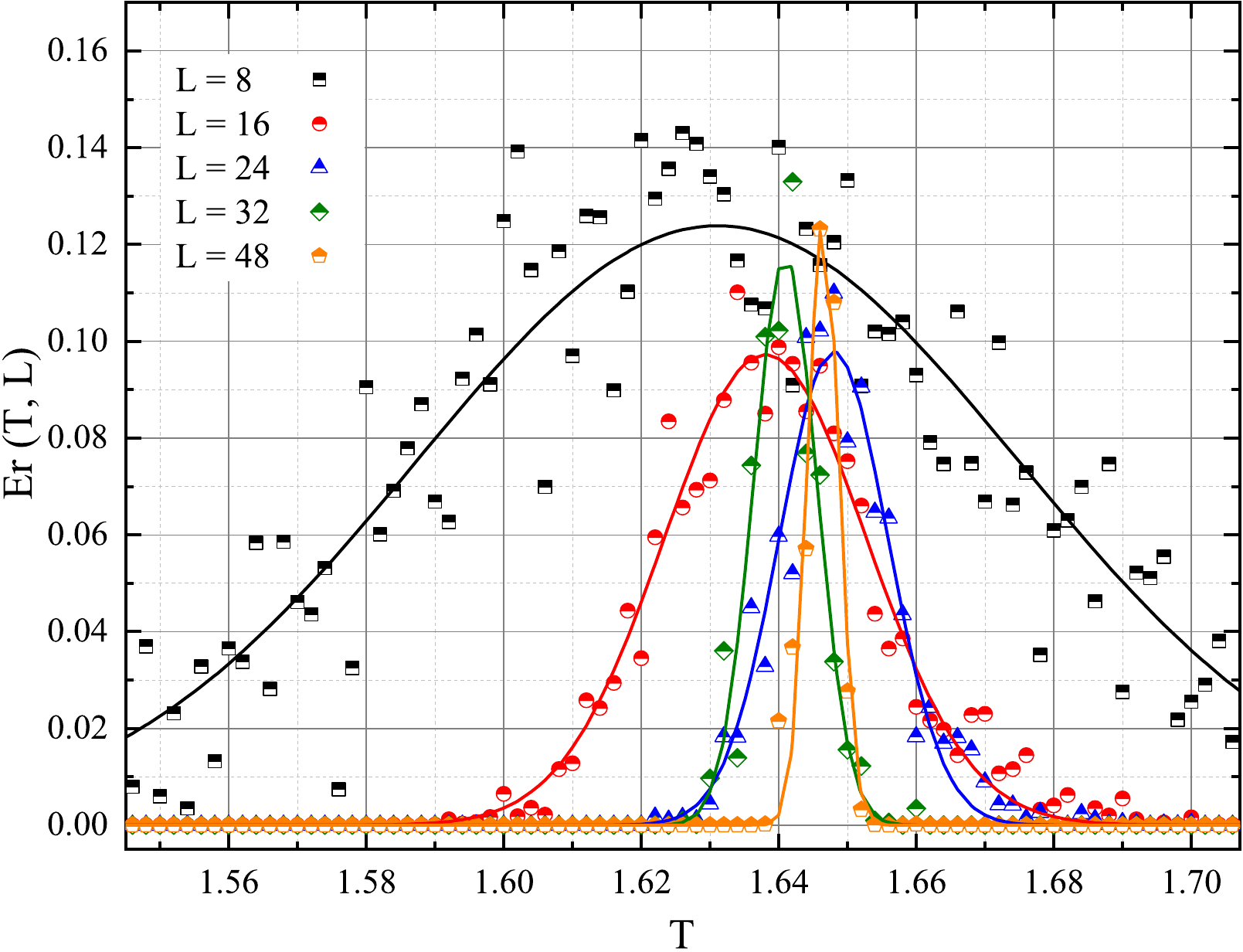}
	\caption{Temperature dependence of the function $\mathrm{Er}(T,L)$. Solid lines are the Gaussian approximation in the critical temperature region.}
	\label{fig:Er}
\end{figure}

Figure \ref{fig:Er} shows the temperature dependence of the function $\mathrm{Er}(T,L)$. A pronounced peak is observed in the vicinity of the critical temperature. The peak has an approximately Gaussian shape, and its width decreases with increasing linear size, which is consistent with finite-size scaling behavior. By approximating each peak with a Gaussian function, we determined the critical temperatures $T_c^{L}$, which were subsequently used to estimate the critical temperature in the thermodynamic limit, $L \to \infty$.

The Gaussian approximation was taken in the form:
\begin{equation}
	\mathrm{Er}(T,L) \sim \exp\left[-\dfrac{\left(T-T_{c}^{L}\right)^2}{{w}^2/{2}}\right]
\end{equation}
where $T_c^{L}$ denotes the peak position and $w$ is the standard deviation.
\begin{figure}
	\centering\includegraphics[scale=0.35]{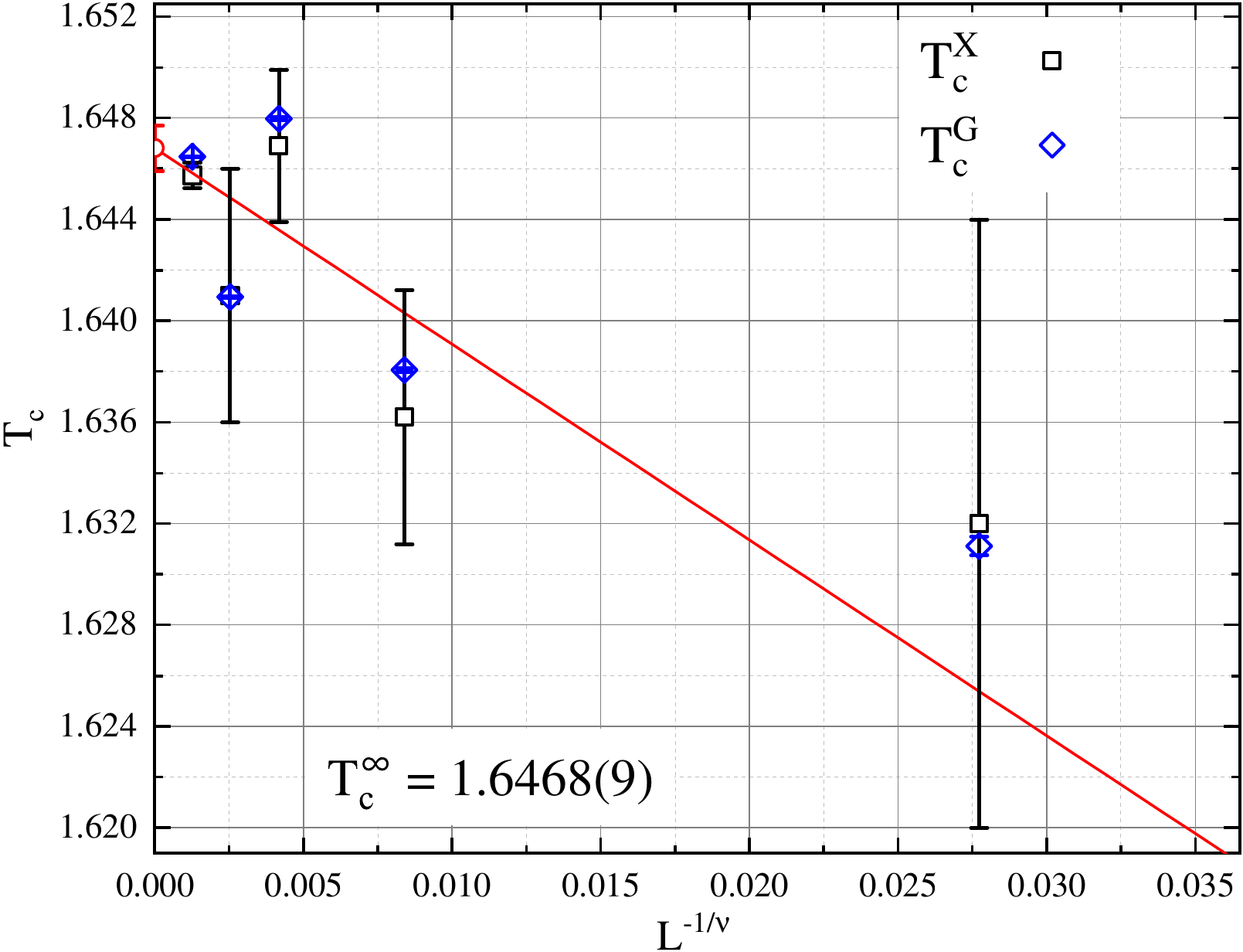}
	\caption{Dependence of critical temperature values on $L^{-1/\nu}$, where $\nu = 0.58$ is the effective critical exponent \cite{Prudnikov_Anis_Heis_Tc}. Determination of $T_c^{\infty}$. }
	\label{fig:Tc_L_1L}
\end{figure}

The critical temperature in the thermodynamic limit, $T_c^{\infty}$, was then determined using a finite-size scaling procedure (see Fig.~\ref{fig:Tc_L_1L}). For this purpose, the critical-temperature estimates $T_c^{X}$ obtained from the crossing of the probability curves (Fig.~\ref{fig:check_P_T}) and $T_c^{G}$ obtained from the Gaussian fits of $\mathrm{Er}(T,L)$ (Fig.~\ref{fig:Er}) were plotted as functions of $L^{-1/\nu}$. 
The finite-size transition temperature was assumed to approach the thermodynamic-limit value
according to 
\begin{equation}
	T_c(L)=T_c^{\infty}+aL^{-1/\nu},
\end{equation}
where $a$ is a nonuniversal constant and $\nu$ is the effective critical exponent for the considered finite-size scaling window. Based on this analysis, the critical temperature was estimated as $T_{c}^{\infty} = \mathbf{1,6468(9)}$.

\begin{figure}
	\centering\includegraphics[scale=0.35]{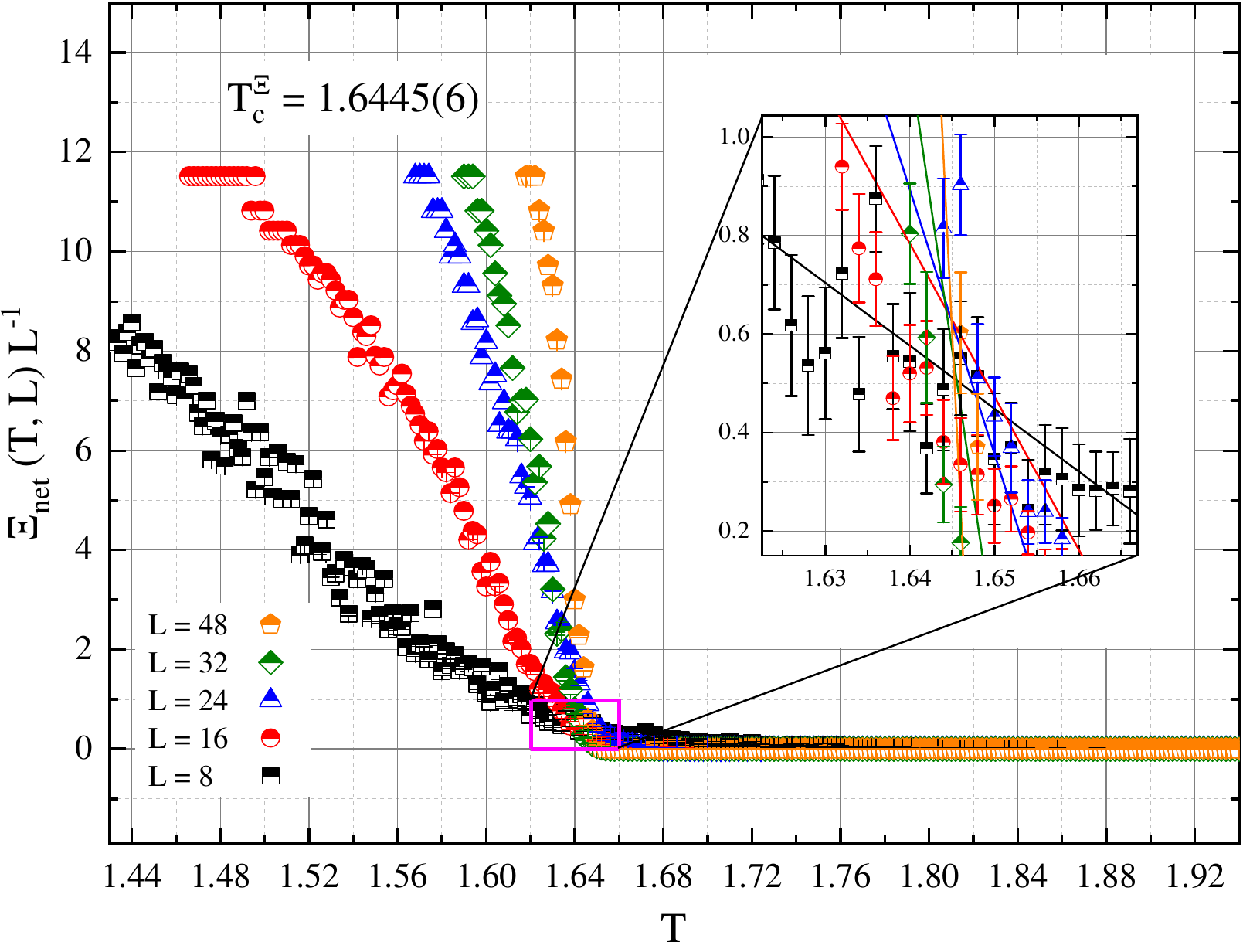}
	\caption{\textbf{Temperature dependence of the CNN-derived finite-size scaling ratio ${\varXi}/{L}$ for the anisotropic Heisenberg model. Finite size analyses lead to the value $T_c^{\varXi}=1.6445(6)$.}}
	\label{fig:ksi_net_heis}
\end{figure}

To calculate \textbf{the correlation-length-like function} $\varXi(T,L)/L$, we used the function defined in Eq.~(\ref{eq:ksi_net}) based on the CNN-based inference of $\mathrm{P_{LT}}(T,L)$. In this framework, the output layer provides a smooth probabilistic observable that can be used to analyze the critical behavior of the system. Figure \ref{fig:ksi_net_heis} shows the temperature dependences of the ratio $\varXi/L$ for different linear sizes.

The critical temperature was determined from the crossing of the corresponding curves for $L=24$, $32$, and $48$. This procedure yields the most accurate estimate obtained in the present study, $T_c^{\varXi} = 1.6445(6)$.

\section{Universal critical scaling}\label{sec:DataAndScaling}
\begin{figure}[h!]
	\centering\includegraphics[scale=0.35]{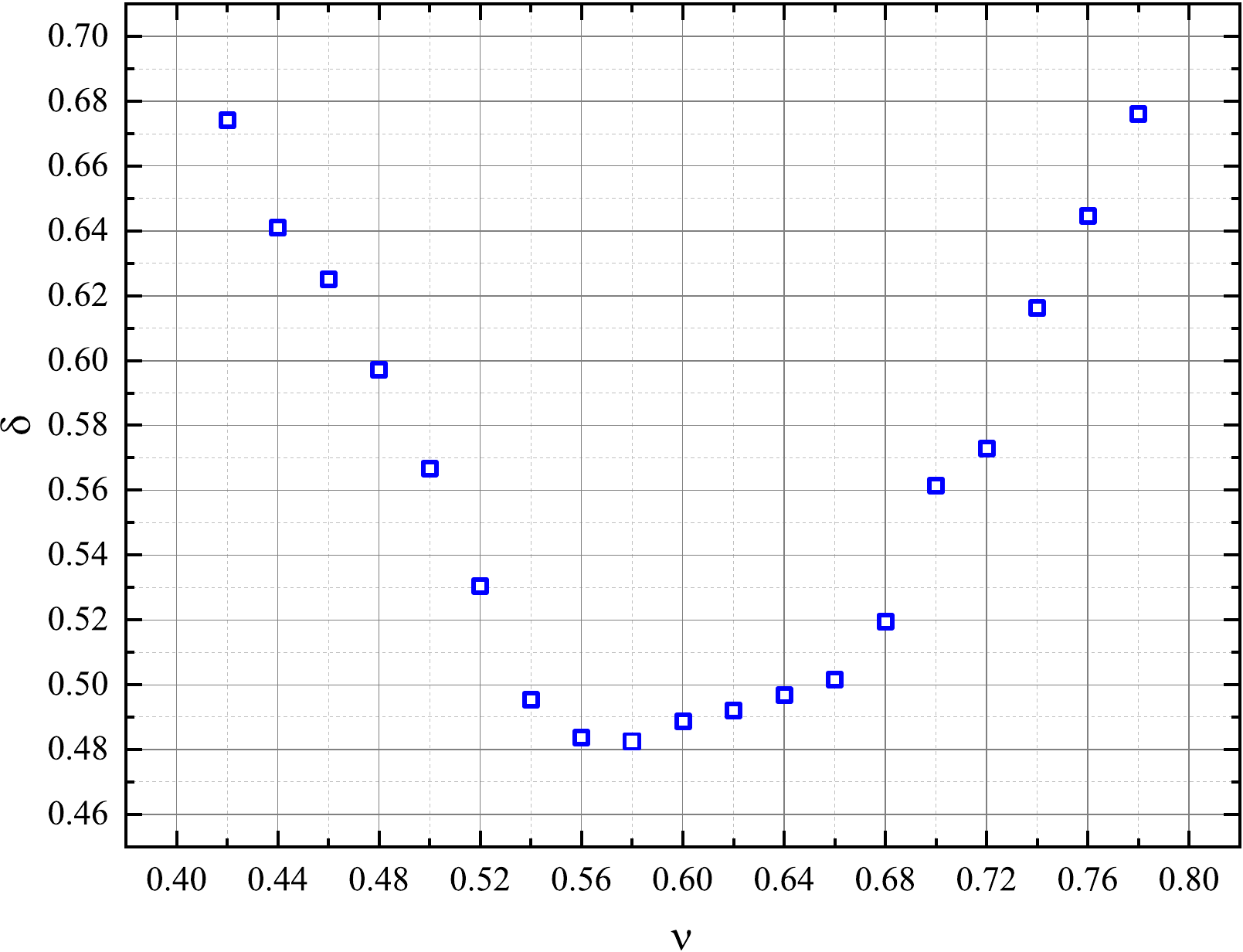}
	\caption{Definition of the critical index $\nu$.}
	\label{fig:check_V}
\end{figure}

To construct the scaling function, one has to determine the value of the critical exponent $\nu$. The scaling function can be expressed as
\begin{equation}
 	\mathrm{P_{\mathbf{LT}}}(T,L) \sim L^{1/v} (T - T_c^X)
\end{equation}
The critical exponent $\nu$ was determined by minimizing the root-mean-square deviation $\delta$ from the optimal scaling collapse shown in Fig.~\ref{fig:check_V}. This procedure yields the estimate $\nu = 0.58(1)$.
It should be emphasized that the value $\nu=0.58(1)$ obtained from the collapse of $\mathrm{P}_{\mathbf{LT}}(T,L)$ is interpreted here as an effective critical exponent for the finite-size scaling window considered in this work. The asymptotic universality class of the three-dimensional easy-axis anisotropic Heisenberg model is the three-dimensional Ising universality class, for which high-precision estimates give $\nu\simeq0.63$. This asymptotic value is expected to be reached in the thermodynamic critical limit, $L\to\infty$, and in the immediate vicinity of the critical point, $\tau=|T-T_c|/T_c\to0$. In the present numerical analysis, however, the collapse is performed for finite lattice sizes $L=8,16,24,32,48$, a finite temperature grid, and a fixed exchange anisotropy	parameter $\Delta=0.63$. Under these conditions, the observed scaling behavior can be affected by finite-size and crossover corrections. Therefore, the exponent extracted from the collapse characterizes an effective scaling regime rather than the final asymptotic fixed point.
\begin{figure}[h!]
	\centering\includegraphics[scale=0.35]{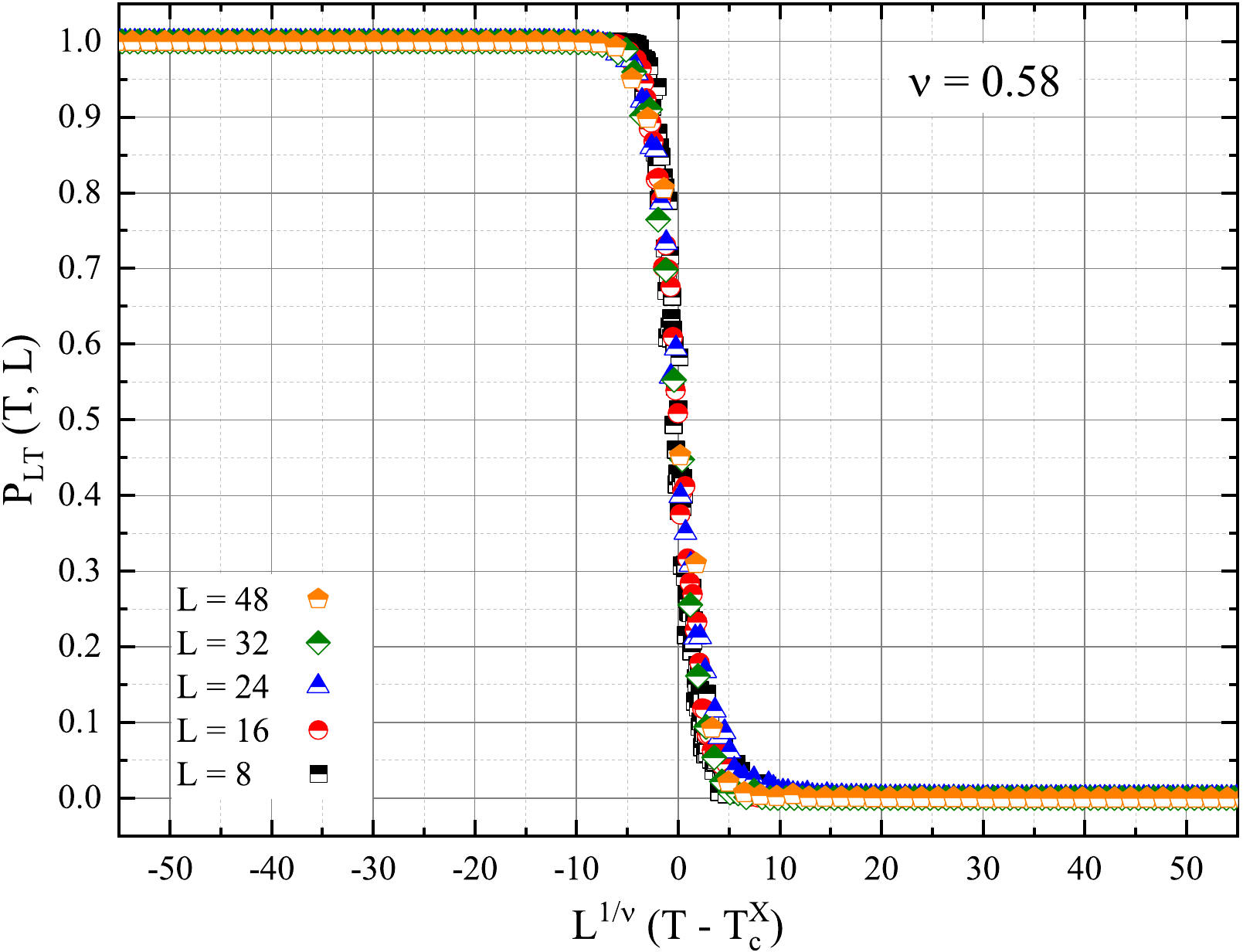}
	\caption{Finite-size scaling collapse of $\mathrm{P_{LT}}(T,L)$ for the anisotropic Heisenberg model for linear sizes $L=8,16,24,32,$ and $48$.
	}
	\label{fig:scaling}
\end{figure}

Figure~\ref{fig:scaling} shows the scaling collapse for the anisotropic three-dimensional Heisenberg model constructed using the obtained exponent $\nu=0.58$. The probability curves $\mathrm{P_{LT}}(T,L)$ for different lattice sizes collapse onto a common scaling dependence in the vicinity of the critical region. This result indicates that the effective value of $\nu$ provides a consistent collapse of the CNN-derived probability curves within the considered finite-size scaling window.

\begin{figure}[h!]
	\centering\includegraphics[scale=0.35]{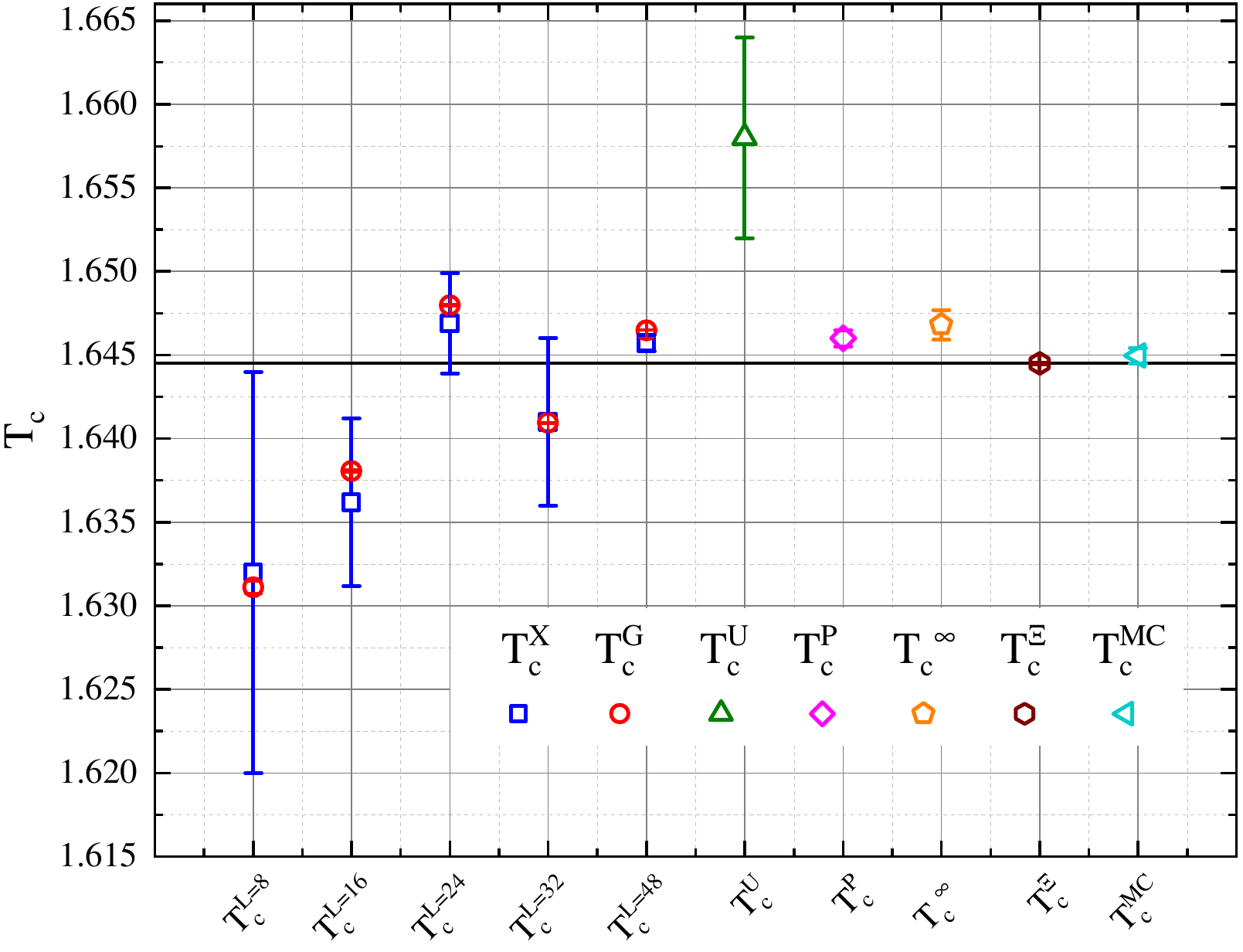}
	\caption{Values of critical temperatures obtained in the study of the 3D Heisenberg model. The solid line corresponds to the value of $T_c^{\varXi}=1.6445(6)$; $T_c^{\mathrm{MC}} = 1.6449(5)$ \cite{Prudnikov_Anis_Heis_Tc}.}
	\label{fig:Tc}
\end{figure}

Figure~\ref{fig:Tc} summarizes the critical-temperature estimates obtained using the different approaches. The values $T_c^X$ and $T_c^G$ show that the uncertainty decreases with increasing linear size, in agreement with the general finite-size behavior expected in Monte Carlo studies. A comparison of the CNN-based estimates $T_c^P$, $T_c^{\infty}$, and $T_c^{\varXi}$ with the Binder-cumulant estimate $T_c^U$ shows that the machine-learning-based procedure yields critical-temperature values that are both accurate and computationally efficient under the same simulation conditions. Among the considered estimators, the value $T_{c}^{\varXi} = 1.6445(6)$ has the smallest deviation and, within statistical uncertainty, is consistent with the reference Monte Carlo result $T_c^{\mathrm{MC}} = 1.6449(5)$ reported in \cite{Prudnikov_Anis_Heis_Tc}.

\section{Conclusion}
\label{sec:concl}
To summarize, in this work we investigated the critical behavior of the anisotropic three-dimensional Heisenberg model with easy-axis anisotropy using a CNN-based approach. The proposed method combines Monte Carlo simulation data, correlation-matrix representations, and CNN-based probabilistic phase classification to determine the critical temperature and analyze the scaling behavior of the system. The results show that this approach provides an efficient and physically meaningful framework for studying second-order phase transitions in anisotropic spin systems.

The most accurate critical-temperature estimate obtained in the present study was $T_c^{\varXi}=1.6445(6)$, which is in very good agreement with the reference Monte Carlo value $T_c^{\mathrm{MC}}=1.6449(5)$. 
In addition, the finite-size scaling analysis yielded the effective critical exponent $\nu=0.58(1)$. This value should be interpreted as an effective exponent characterizing the considered finite-size scaling window at the fixed anisotropy parameter $\Delta=0.63$, rather than as a replacement for the asymptotic three-dimensional Ising exponent. The corresponding scaling collapse demonstrates the consistency of the CNN-derived probability curves within the finite-size range considered in this work.

An important advantage of the proposed machine-learning approach is that it allows the critical temperature to be determined in several complementary ways from the same CNN-based inference. In particular, $T_c$ was estimated from the crossing of the probability curves $P_{LT}(T,L)$, from the peak positions of the variance-like function $Er(T,L)$, and from the finite-size behavior of the ratio $\varXi(T,L)/L$. The possibility of obtaining the critical temperature by several independent procedures within the same computational framework is a significant benefit, since it provides internal consistency checks and increases the robustness of the analysis. 
Among these estimators, the value obtained from the CNN-derived finite-size scaling observable $\varXi(T,L)/L$, $T_c^{\varXi}$, showed the best agreement with the reference Monte Carlo result.

Another important result of this work is the computational efficiency of the CNN-based procedure. Under the same simulation conditions, the calculation of correlation matrix and their subsequent machine-learning analysis required substantially less computational time than the corresponding classical thermodynamic analysis based on Binder cumulants. For the largest system considered, this reduction reached nearly an order of magnitude, while the resulting critical-temperature estimates remained consistent with the Monte Carlo reference value. At the same time, the present study shows that the accuracy of the method depends strongly on the density of the temperature grid in the critical region, so that sufficiently fine temperature sampling is essential for reliable determination of the transition point.

Overall, the results demonstrate that CNNs can serve not only as classification tools but also as effective auxiliary instruments for extracting thermodynamically relevant quantities from simulated spin configurations. The proposed approach provides a practical route to the analysis of critical phenomena, combines computational efficiency with several independent estimators of $T_c$, and opens further perspectives for the application of machine-learning methods in condensed-matter physics.

\ack{This work was supported by the Ministry of Science and Higher Education of the Russian Federation within the governmental assignment for Boreskov Institute of Catalysis (project FWUR-2024-0039).}

\roles{Alina A. Chubarova: Methodology, Software, Writing -- original draft. Ivan A. Mamonov: Data curation. Marina V. Mamonova: Project administration. Mikhail I. Bogachev: Conceptualization, Writing -- review and editing. Pavel V. Prudnikov: Conceptualization, Supervision, Writing -- original draft}

\data{The data that support the findings of this study are available from the corresponding author upon reasonable request.}

\suppdata{No supplementary material is associated with this article.}

\end{document}